\documentclass[conference]{IEEEtran}
\IEEEoverridecommandlockouts

\usepackage{cite}
\usepackage{amsmath,amssymb,amsfonts}
\usepackage{algorithmic}
\usepackage{graphicx}
\usepackage{textcomp}
\usepackage{xcolor}
\usepackage{graphicx} 
\usepackage{algorithm}
\usepackage{algorithmic}
\usepackage{caption}
\usepackage{subfigure}
\usepackage{rotfloat}
\usepackage{multirow}
\usepackage{colortbl}
\usepackage{graphicx}
\usepackage{ulem}
\usepackage{pifont}
\usepackage{url}
\usepackage{booktabs}
\usepackage{threeparttable}

\usepackage{hyperref}

\def\BibTeX{{\rm B\kern-.05em{\sc i\kern-.025em b}\kern-.08em
    T\kern-.1667em\lower.7ex\hbox{E}\kern-.125emX}}

\begin{document}
\title{FastCHGNet: Training one Universal Interatomic Potential to 1.5 Hours with 32 GPUs}

\author{
    Yuanchang Zhou$^{1,2}$, Siyu Hu$^{1,2,*}$, Chen Wang$^{1,2}$, Lin-Wang Wang$^{2,3}$, Guangming Tan$^{1,2}$, Weile Jia$^{1,2,*}$ \\
    \IEEEauthorblockA{
        $^{1}$State Key Lab of Processors, Institute of Computing Technology, Chinese Academy of Sciences \\
        $^{2}$University of Chinese Academy of Sciences \\
        $^{3}$Institute of Semiconductor, Chinese Academy of Sciences \\
        Email: \{zhouyuanchang23s, husiyu, tgm, jiaweile\}@ict.ac.cn, wangchen246@mails.ucas.ac.cn, lwwang@semi.ac.cn
    }
    \thanks{*Corresponding author}
}


\maketitle

\begin{abstract}

Graph neural network universal interatomic potentials (GNN-UIPs) have demonstrated remarkable generalization and transfer capabilities in material discovery and property prediction. These models can accelerate molecular dynamics (MD) simulation by several orders of magnitude while maintaining \textit{ab initio} accuracy, making them a promising new paradigm in material simulations. One notable example is Crystal Hamiltonian Graph Neural Network (CHGNet), pretrained on the energies, forces, stresses, and magnetic moments from the MPtrj dataset, representing a state-of-the-art GNN-UIP model for charge-informed MD simulations. 
However, training the CHGNet model is time-consuming(8.3 days on one A100 GPU) for three reasons: (i) requiring multi-layer propagation to reach more distant atom information, (ii) requiring second-order derivatives calculation to finish weights updating and (iii) the implementation of reference CHGNet does not fully leverage the computational capabilities. This paper introduces FastCHGNet, an optimized CHGNet, with three contributions: Firstly, we design innovative Force/Stress Readout modules to decompose Force/Stress prediction. Secondly, we adopt massive optimizations such as kernel fusion, redundancy bypass, etc, to exploit GPU computation power sufficiently. Finally, we extend CHGNet to support multiple GPUs and propose a load-balancing technique to enhance GPU utilization. Numerical results show that FastCHGNet reduces memory footprint by a factor of 3.59. The final training time of FastCHGNet can be decreased to \textbf{1.53 hours} on 32 GPUs without sacrificing model accuracy.

\end{abstract}


\section{Introduction}

Atomic-level simulations based on Density Functional Theory (DFT) calculations have significantly advanced materials modeling over the past few decades. Recently, the emergence of Universal Interatomic Potentials (UIP) has opened new opportunities for modeling complex materials, such as alloys, amorphous solids, condensed phase liquids, and nanostructured materials~\cite{park2024scalable, batatia2023foundation, deng2023chgnet}. 
Unlike dedicated interatomic potentials like DeePMD-kit~\cite{wang2018deepmd}, DTNN~\cite{schutt2017quantum}, SchNet~\cite{schutt2017schnet}, HIP-NN~\cite{lubbers2018hierarchical}, PhysNet~\cite{unke2019physnet}, or DimeNet~\cite{gasteiger_dimenet_2020} that train a separate model for each individual physical system,
UIP models are trained on extensive DFT datasets that cover a wide range of elements, aiming to capture the fundamental physics of atomic interactions.
Once trained, a UIP model can be applied to diverse physical systems without requiring further  DFT calculations. For example, 
Crystal Hamiltonian Graph Neural Network (CHGNet)~\cite{deng2023chgnet} and MACE~\cite{batatia2022mace, batatia2023foundation} have demonstrated remarkable generalization and transfer capabilities. While maintaining exceptional accuracy, UIP models continue to show great potential in materials science and chemistry, significantly advancing our understanding of atomic interactions and complex materials behavior.

One state-of-the-art UIP model is CHGNet. It is currently the only charge-informed Graph Neural Network(GNN) based interatomic potential (trained with the magnetic moments) and has demonstrated excellent results on various Lithium battery related physical systems. 
Based on its predicted magnetic moments, CHGNet can accurately represent the orbital occupancy of electrons. It presents strong reliability in predicting properties such as conductivity and activation energies across various structures and compositions (e.g., Li$_x$FePO$_4$, LiMnO$_2$)~\cite{deng2023chgnet}.

Despite the remarkable capabilities of CHGNet, its training time is still a significant bottleneck. Training CHGNet on the Materials Project Trajectory Dataset (MPtrj, with 1,580,395 atom configurations) using a single A100 GPU takes about 8.3 days.
This extended training time limits the ability to iterate and improve the model quickly, making it difficult for designing new UIP models. 
We provide an in-depth examination and find that there are three main reasons for the long training time of CHGNet. 
\textit{First}, the complex model architecture. CHGNet is a GNN-based model and it implements a complex forward pass. Each central atom acquires information from neighboring atoms to update its embedding. By increasing the GNN layers, a central atom can interact with more distant neighbors. For example, in CHGNet, the average number of neighboring atoms increases exponentially with the number of interaction blocks, reaching 104, 10,795, and 1,121,797 for interaction blocks 1 to 3. 
In practice, the number of interaction blocks often exceeds 3, meaning each atom update can involve over 1 million neighbors. Additionally, the number of bonds and angulars also increases exponentially. This complex model architecture results in heavy computational costs, leading to extensive training times. 
\textit{Second}, calculating second-order derivatives is necessary for updating model weights. CHGNet predicts energy, force, stress and magnetic moment. According to conservation laws, force is derived by differentiating total energy with respect to atomic positions, while stress is derived by differentiating total energy with respect to the lattice strain tensor.
As a result, second-order derivatives are required when using the Adam optimizer to update model weights. The second-order derivatives calculation is a time-consuming process due to the high computational complexity. 
\textit{Third}, the implementation of reference CHGNet is inefficient. The original CHGNet can only be trained on single GPU with a small minibatch size. This reference implementation contains many serial operations that are not efficiently parallelized. Also, numerous redundant computations that have not been carefully examined and eliminated are included in the implementation of reference CHGNet. It has exhibited high memory usage and large launched CPU kernels in the training. 
On the whole, there is a lot of room for further optimization and efficiency improvements in the overall training process of CHGNet.

In this paper, we propose FastCHGNet, a highly optimized version of CHGNet. We design a series of strategies to enhance the training efficiency of reference CHGNet. Our major contributions are:
\begin{itemize}
    \item We propose an innovative module to decouple Energy-Force and Energy-Stress. Meanwhile, we give strict proof to prove that the Force decomposition module meets the rotation equivariant principle.
    \item We perform efficient parallel optimization strategies such as batching, kernel fusion, redundancy elimination, computational results reuse, etc, to saturate the GPU computation resources.
    \item We implement an efficient large batch training process by leveraging multi-GPUs. The training parameters are heuristically tuned to ensure a steady and reliable convergence. The distribution of atoms, bonds, and angles is also carefully considered to perform load balance.
    \item With no sacrifice of accuracy, the training time of CHGNet (8.3 days) can be reduced to 1.53 hours (by using 32 NVIDIA A100 GPUs), gaining a 130x speedup. 
\end{itemize}

\section{Background}
GNNs have been widely investigated for modeling many-body interactions and have led to a revolution in molecule and atomistic modeling. We will give the general formalization of the message-passing mechanism. This is followed by a rigorous expression of CHGNet workflow.

\subsection{Massage Passing Mechanism}
Consider a graph $\mathcal{G} = \left\{ \mathcal{V}, \mathcal{E} \right\} $, where $\mathcal{V}$ represents the set of vertices and $\mathcal{E}$ represents the set of edges. Each atom $i$ can be viewed as a node and atom feature at layer $t$ denoted by $h_i^t$. The interaction of atom $i$ and neighbor atom $j$ within a fixed cutoff distance can be regarded as an edge, denoted by $e_{ij}$. As shown in Eq.~\ref{eq: aggregate and update}, GNN propagates information across neighboring edges and aggregates the information into the central atom representation. A non-linear node update operation is applied after the aggregation, where $M_t$ and $U_t$ are learnable messages and node update functions respectively. The central atom can receive the further neighbor atoms' information by stacking GNN interaction blocks. 

\begin{equation}
	\begin{split}
        \mathbf{m}_i^{t+1} &= \sum_{j \in \mathcal{N}(i)} M_t \left( \mathbf{h}_i^t, 
        \mathbf{h}_j^t, \mathbf{e}_{ij} \right) \\
        \mathbf{h}_i^{t+1} &= {U}_t \left( \mathbf{h}_i^t, \mathbf{m}_i^{t+1} \right)
    \end{split}
    \tag{1}
    \label{eq: aggregate and update}
\end{equation}

\begin{figure*}[t]
    \centering
    \includegraphics[width=0.8\linewidth]{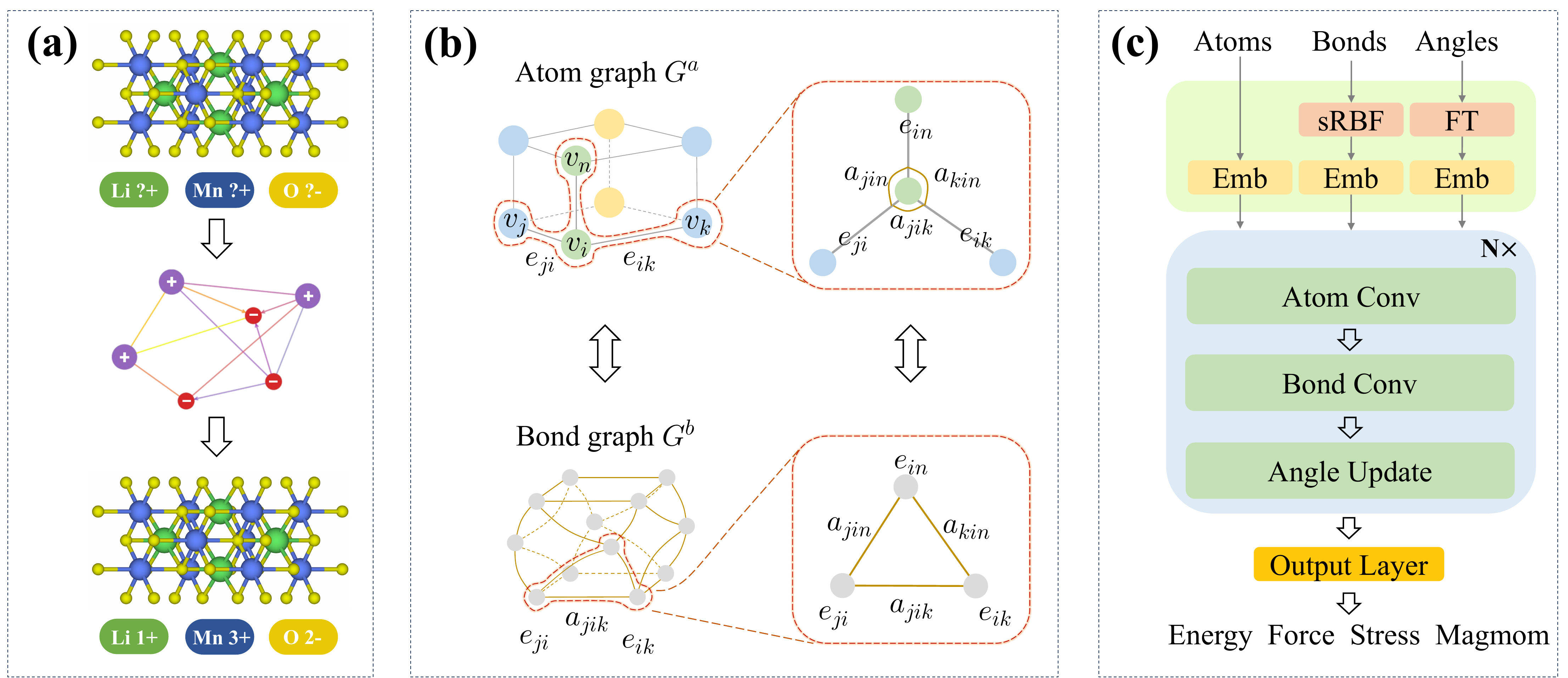}
    \caption{The CHGNet workflow. (a) The input and output of CHGNet; (b)The graph representation of crystal structure; (c) The core module of CHGNet.}
    \label{fig: CHGNet workflow}
\end{figure*}
\subsection{CHGNet}
CHGNet is pre-trained on the Materials Project Trajectory Dataset with about 1.5 million inorganic structures under Density Functional Theory (DFT) calculation. The workflow of CHGNet is depicted in Fig.~\ref{fig: CHGNet workflow}(a). CHGNet accepts crystal structures with undetermined atomic charges and predicts corresponding energy, forces, stress, and magnetic moments. Based on the predicted magnetic moments, atomic charges can be inferred under inherent charge constraints. The CHGNet can be divided into 4 parts, Molecular Graph Extraction, Feature Embeddings, Interaction Block, and Output Layer. The details of each part are described as follows.

(1)\textbf{Molecular Graph Extraction:} Given a crystal structure with periodic boundary conditions, an atom graph $G^a$ and an auxiliary bond graph $G^b$ can be produced, shown in Fig.~\ref{fig: CHGNet workflow}(b). (a) $G^a$ is used to represent two-body interaction. In $G^a$, each atom $i$ is treated as a node $v_i$. Each atom identifies neighboring atoms within its cutoff radius, and the edge $e_{ij}$ connecting node $v_i$ and node $v_j$ is initialized by Cartesian distance $r_{ij}:=| r_j - r_i| $. (b) $G^b$ is used to represent three-body interaction. The node in $G^b$ reuses the edge representation in $G^a$. The edge $a_{ijk}$ in $G^b$ is indicated by angle $\theta_{ijk} = arccos \frac{r_{ij} \cdot r_{ik}}{|r_{ij}| \cdot |r_{ik}| } $ for pairwise information between $e_{ij}$ and $e_{ik}$ in $G^a$. The number of atoms, bonds, and angles in the crystal structure is written as $N_v$, $N_b$ and $N_a$ respectively.

(2)\textbf{Feature Embedding:} A linear transformation is applied to atomic numbers to get the initialized node feature, denoted as $v^0_i$. The distance $r_{ij}$ and angular $\theta_{ijk}$ are first expanded by trainable smooth Radial Bessel Function(noted as sRBF)\cite{gasteiger_dimenet_2020} and Fourier Transformation(denoted as $\mathrm{FT}$) respectively, and linear transformations($\mathcal{L}(x) = x \mathbf{W} + b$) are then applied to get the bond feature $e^0_{ij}$, bond weights in atom-conv module $e^a_{ij}$, bond weights in bond-conv module $e^b_{ij}$, and angle feature $a^0_{ijk}$. The transformations are defined as follows:
\begin{equation*}
\begin{split}
    v^0_i &= \mathrm{Z}_i \mathbf{W}_v \colon \mathbb{R}^{N_v \times 1} \rightarrow  \mathbb{R}^{N_v \times 64} \\
    [e^0_{ij}, e^a_{ij}, e^b_{ij}] &= \mathcal{L} \big( \mathrm{sRBF}(r_{ij})) \colon \mathbb{R}^{N_v \times 1} \rightarrow  \mathbb{R}^{2N_b \times (3\times 64)} \\
    a^0_{ijk} &= \mathcal{L} \big( \mathrm{FT}(\theta_{ijk}) \big) \colon \mathbb{R}^{N_a \times 1} \rightarrow  \mathbb{R}^{N_a \times 64} \\ 
\end{split}
\tag{2}
\end{equation*}

(3)\textbf{Interaction Block:} The essential parts of CHGNet model are illustrated in Fig.~\ref{fig: CHGNet workflow}(c). The interaction block is in the box with blue background which encodes and updates atoms, bonds, and angles embeddings upon the pre-defined $G^a$ and $G^b$. The interaction block in the $t$-th layers:
\begin{equation*}
    \begin{split}
    IB^t \colon [v^t_i, e^t_{ij}, a^t_{ijk}, e^a_{ij}, e^b_{ij} ] \rightarrow [v^{t+1}_i, e^{t+1}_{ij}, a^{t+1}_{ijk} ]  
    \end{split}
    \tag{3}
\end{equation*}
where $t \in 0,1,2$ is the layer of the interaction block. The interaction block contains:
\begin{itemize}
    \item Atom Conv: node feature and pairwise bond feature are concatenated $f_v = [v^t_i, v^t_j, e^t_{ij}]$ and a GatedMLP $\phi^t_v$ is performed on $f_v$. The weighted message is constructed and aggregated in the following form:   
    \begin{equation*}
        \begin{split}
        v^{t+1}_i = v^t_i + \mathcal{L}^t_v [ \sum_{j \in N(i)} e^a_{ij} \odot \phi^t_v(f_v) ] 
        \end{split}
        \tag{4}
    \end{equation*}
    \item Bond Conv: the input of the bond convolution module is $f_e = [v^{t+1}_i, e^t_{ij}, e^t_{ik}, a^t_{ijk}]$ and a GatedMLP $\phi^t_e$ is applied on $f_e$. The bond feature is updated by:
    \begin{equation*}
    \begin{split}
        e^{t+1}_{ij} = e^t_{ij} + \mathcal{L}^t_e [ \sum_{k \in N(i), k \neq j} e^b_{ij} \odot e^b_{ik} \odot \phi^t_e(f_e) ]
    \end{split}
    \tag{5}
    \end{equation*}
    \item Angle Update: The input of the Angle Update module is the renewed node, and bond feature, combined with the stale angle features, $f_a = [v^{t+1}_i, e^{t+1}_{ij}, e^{t+1}_{ik}, a^t_{ijk}]$. The angle feature is calculated by:
    \begin{equation*}
    \begin{split}
        a^{t+1}_{ijk} = a^t_{ijk} + \phi^t_a(f_a)
    \end{split}
    \tag{6}
    \end{equation*}
\end{itemize}

where GatedMLP operation\cite{xie2018crystal} is formulated by $\phi(x) = (\sigma \circ \mathrm{LN} \circ \mathrm{Fc}(x)) \odot (\mathrm{g} \circ \mathrm{LN} \circ \mathrm{Fc}(x))$. $\sigma$ and $\mathrm{g}$ are Sigmoid and SiLu activations. $\mathrm{LN}$ and $\mathrm{Fc}$ are represented by layer normalization and fully connected operation. $\odot$ is elementwise multiplication.

(4)\textbf{Output Layer:} The total energy is derived by summing up the nonlinear projections of the final atomic features. Forces and stress are calculated by automatically differentiating the energy for atomic coordinates and the lattice strain tensor.

\section{Method}

The CHGNet introduced in the Background suffers from long training time. We develop a set of strategies to fully exploit the computing power of the heterogeneous architecture. Various optimizations for CHGNet are implemented, including output layer decomposition, kernel fusion, redundancy bypass, load balance, etc. The optimized CHGNet is called FastCHGNet.


\subsection{Overview of FastCHGNet}
\begin{figure*}
    \centering
    \includegraphics[width=\linewidth]{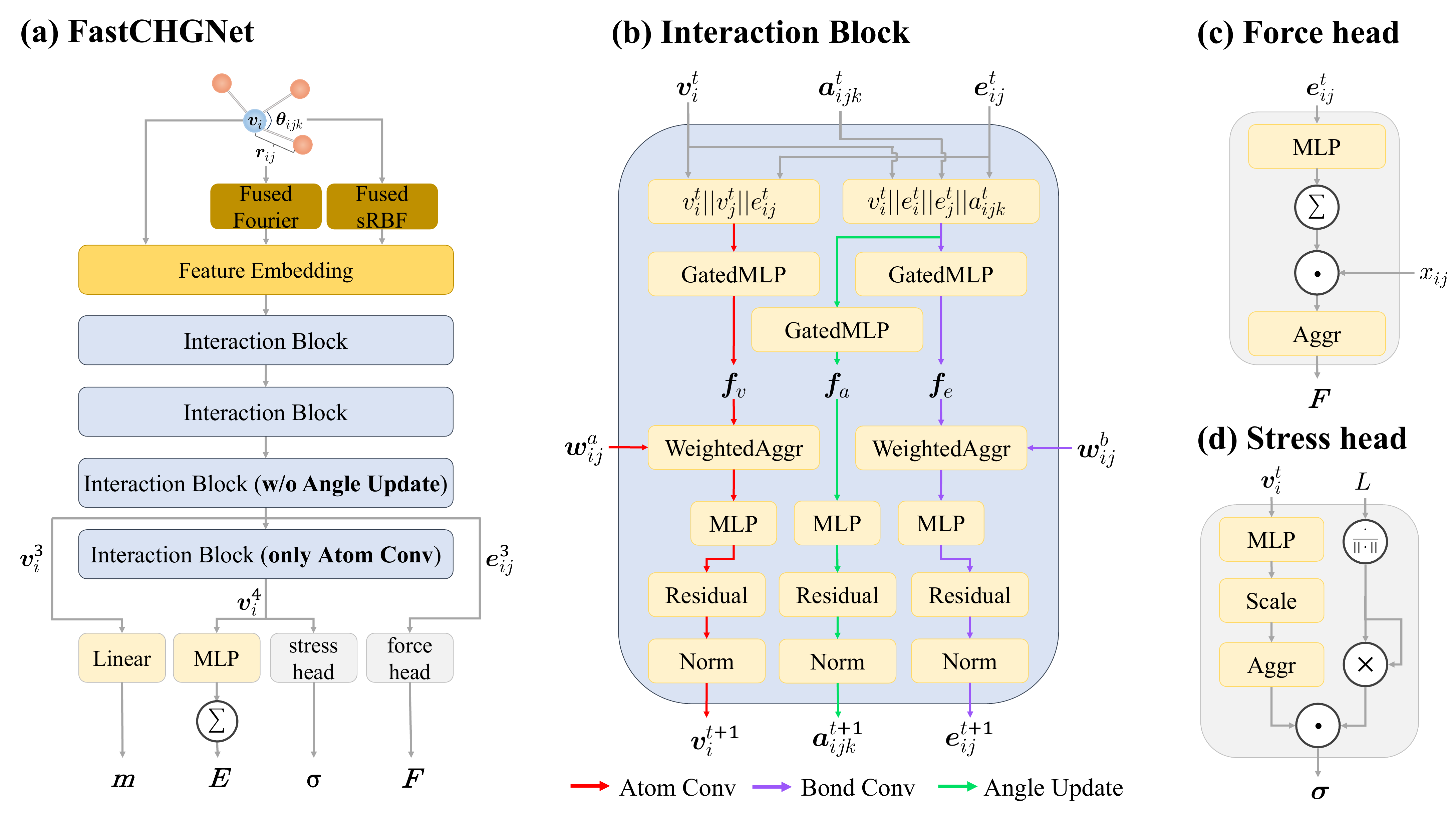}
    \caption{The architecture of FastCHGNet. 
    (a) the high-level workflow of FastCHGNet;
    (b) the detailed implementation of Interaction Block;
    (c) Force head module;
    (d) Stress head module.
    }
    \label{fig: FastCHGNet arch}
\end{figure*}

The structure of FastCHGNet is shown in Fig.~\ref{fig: FastCHGNet arch}(a). We will introduce FastCHGNet from top to bottom. The referenced `sRBF' and `Fourier' modules contain numerous element-wise operations. We perform aggressive kernel fusion. The optimized modules are denoted as `Fused-sRBF' and `Fused-Fourier' modules. The Feature Embedding module remains unchanged. The Interaction Block of FastCHGNet is much more efficient than the original Interaction Block because of the dependency elimination of the bond features and the angle features. Besides, multiple scattered small kernels have been fused and the redundant computation has been removed in the Interaction Block. In the prediction of Energy/Force/Stress, we design a novel multi-head module to decouple these properties instead of calculating Force/Stress through the rigorous derivative. 

We categorize the optimizations into two major classes, `Model innovation' and `System optimizations'. The multi-head decoupling and dependency elimination are introduced in the `Model innovation'. The kernel fusion, load balance, etc are detailed in `System Optimizations'.

\subsection{Model innovation}
\textbf{Multi-Head Decomposition:} In the reference CHGNet, the Force and Stress are calculated by $F_i = -\frac{\partial E_{\text{tot}}}{\partial x_i}$ and $\sigma = \frac{1}{V} \frac{\partial E_{\text{tot}}}{\partial \epsilon}$, under the guidance of physical conservative rule. The first-order derivatives are required to calculate Force and Stress. The second-order derivatives are needed to update weights. Note that second-order derivatives calculation has higher computational complexity which is storage and computation-intensive.

However, this physical rule is proven not compulsory in tasks such as the initial structure to relaxed energy (IS2RE). Predicting $F_i$ directly can yield better fitting results~\cite{gasteiger2021gemnet}. In FastCHGNet, we design a sound Force Head and a Stress Head to fit Force and Stress directly. We provide rigorous proof to claim that our designed Force decomposition module assures an important property: rotation equivariance.

Force Head: The Force Head in FastCHGNet is shown in Fig.~\ref{fig: FastCHGNet arch}(c). This head directly predicts the force using the final bond features $e^t_{ij}$ combined with the bond vectors ${x}_{ij}$. It is defined as follows:
\begin{equation*}
	\begin{split}
        {n}_{ij} &= MLP(e^t_{ij}) \\
        F_i &= \sum_j(n_{ij} \odot {x}_{ij})
        \end{split}
        \tag{7}
\end{equation*}
where $n_{ij}$ and ${x}_{ij}$ represent the magnitude and direction of the force exerted by atom j on atom i respectively. The net force $F_i$ is the sum of the forces exerted on i by all neighboring atoms j, which is represented by $n_{ij} \odot {x}_{ij}$.
For any rotation matrix $R$, the bond features $e^t_{ij}$ are invariant and the bond vectors ${x}_{ij}$ is transformed into $R{x}_{ij}$. Thus, the Force head is rotation equivariant as shown below:
\begin{equation*}
	\begin{split}
        F'_i&=\sum_j(n_{ij} \odot R{x}_{ij})\\
        &=R\sum_j(n_{ij} \odot {x}_{ij}) \\
        &=RF_i
        \end{split}
        \tag{8}
\end{equation*}

Stress Head: Similarly, we compute the lattice's cross-product to obtain the normal vector of the lattice, and combine it with the atomic representation 
$v^t_i$ from the final layer to calculate the system's stress. The Stress Head is defined as follows: 
\begin{equation*}
	\begin{split}
        \sigma = \sum_i( (scale*MLP(v^t_i)) \odot (\sum_{ij}\frac{L_i}{||L_i||}\otimes\frac{L_j}{||L_j||}) )
    \end{split}
    \tag{9}
\end{equation*}
where $v^t_i$ is the last layer atom feature and $L$ is lattice vectors. $\otimes$ denotes outer products. The structure is depicted in Fig.~\ref{fig: FastCHGNet arch}(d).

\textbf{Dependency Elimination:} In CHGNet, message construction is based on atom, bond, and angle features, which are updated in each Interaction Block. In the Atom Convolution, messages are derived from the concatenated feature vectors of an atom and a bond, specifically $v^t_i$ and $e^t_{ij}$. For the Bond Convolution, messages come from the updated atomic features $v^{t+1}_i$, the bond features $e^t_{ij}$ and the angle features $a^t_{ijk}$. In the Angle Update layer, the messages used to update the angle representation are obtained from the updated atomic features $v^{t+1}_i$ and the bond features $e^{t+1}_{ij}$, along with the angle features $a^t_{ijk}$. In the reference Interaction Block, the dependencies are as follows:
\begin{equation*}
	\begin{split}
       v^{t+1}_i &= \text{Atom Conv}(v^t_i, e^t_{ij}) \\
       e^{t+1}_{ij} &= \text{Bond Conv}(v^{t+1}_{i}, e^t_{ij}, a^t_{ijk}) \\
       a^{t+1}_i &= \text{Angle Update}(v^{t+1}_i, e^{t+1}_{ij}, a^t_{ijk})
       \end{split}
       \tag{10}
\end{equation*}

In FastCHGNet, as shown in Fig.~\ref{fig: FastCHGNet arch}(b), 
we break the dependency of  $e^{t+1}_{ij}$ and $a^{t+1}_i$, directly utilizing the feature vectors $v^{t}_i$, $e^t_{ij}$ to construct the messages in the bond convolution layer and angle update layer. The dependency of Atom Conv module, Bond Conv module, and Angle Update module is eliminated. The inputs for Bond Conv module and Angle Update module become consistent. Both of the inputs require going through the GatedMLP operation. 
\begin{equation*}
	\begin{split}
       e^{t+1}_{ij}&=\text{Bond Conv}(v^{t}_{i}, e^t_{ij}, a^t_{ijk}) \\
       a^{t+1}_i&=\text{Angle Update}(v^{t}_i, e^{t}_{ij}, a^t_{ijk})
       \end{split}
       \tag{11}
\end{equation*}

This modification does not affect accuracy while allowing the forward pass for atoms, bonds, and angles to update concurrently.

\subsection{System optimizations}
\begin{figure}
    \centering
    \includegraphics[width=\linewidth]{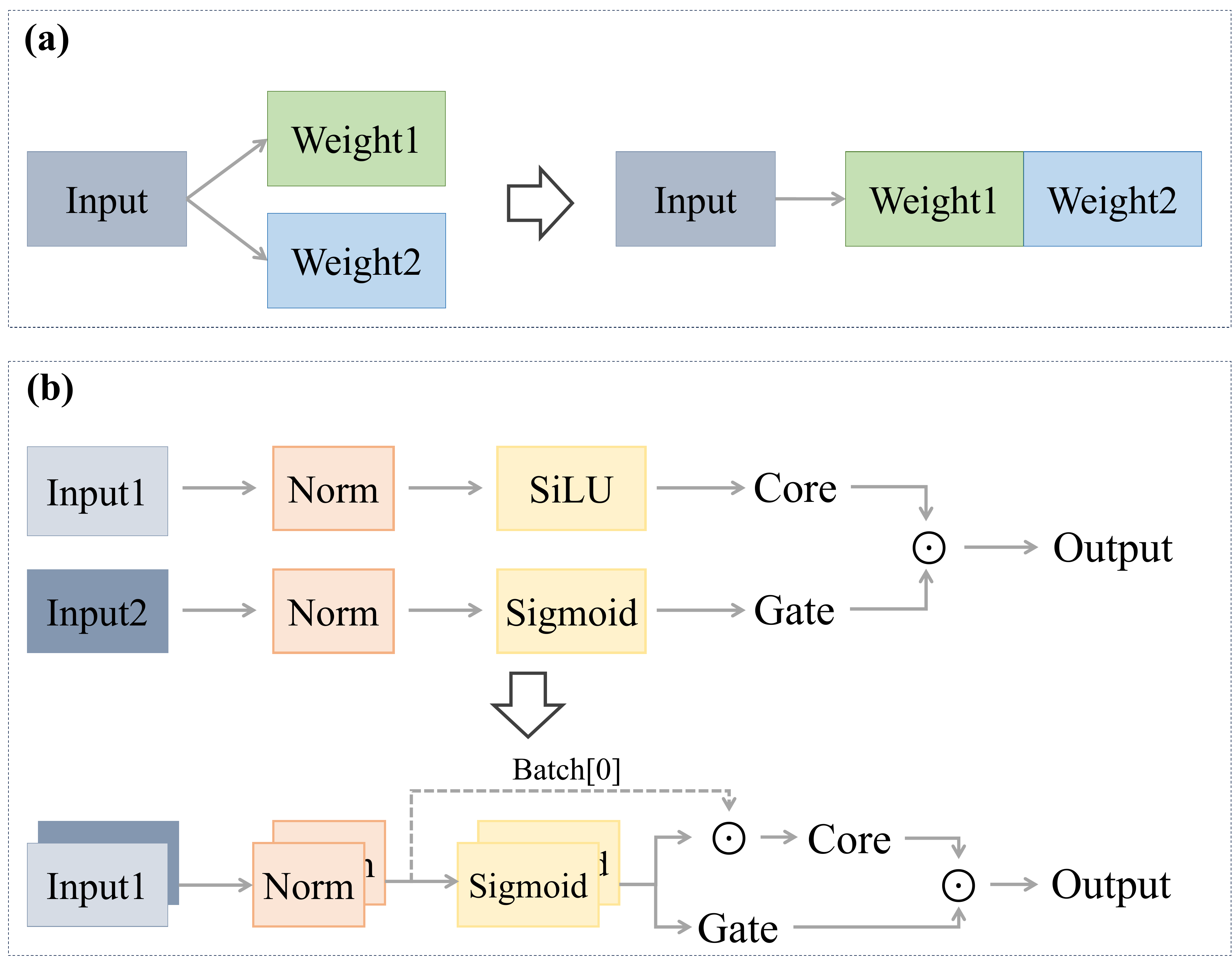}
    \caption{Packing strategy. (a) Sharing the same input can be fused into a larger matrix-matrix multiplication;  
    (b) The core branch and gate branch fusion strategy of GatedMLP operation. 
    }
    \label{fig: Packing strategy}
\end{figure}
\textbf{Parallel Computation of Basis:} As depicted in Alg.~\ref{alg: serial process}, CHGNet implements the serial computation of $\tilde{e}^a_{ij}$, $\tilde{e}^b_{ij}$, ${\tilde{a}_{ijk}}$, for each sample in one batch. It iterates through each graph $g$ to extract the lattice $L \in \mathbb{R}^{3\times3}$ and the neighbor image $I \in \mathbb{R}^{n\times3}$, subsequently computing $r_{card}$ along with the Cartesian coordinates of the starting atom $r_i$ and the destination atom $r_j$. '@' denotes matrix multiplication. The algorithm then employs the smooth radial basis function $sRBF$ and fourier transformation $FT$ to compute $g\_{\tilde{e}^a_{ij}}$, $g\_{\tilde{e}^b_{ij}}$ and $g\_{\tilde{a}_{ijk}}$, appending these values to their respective lists. Ultimately, the results are concatenated along dimension 0 and returned.
However, a significant limitation of this algorithm is its reliance on a serial processing approach, which restricts the effective utilization of the parallel capabilities of modern computational hardware. This sequential computation incurs considerable CPU overhead, especially when managing large batch sizes, ultimately resulting in decreased overall performance.

\begin{algorithm}[t]
    \caption{Serial Computation of Basis in Batches}
    \label{alg: serial process}
    \renewcommand{\algorithmicrequire}{\textbf{Input:}}
    \renewcommand{\algorithmicensure}{\textbf{Output:}}
    \begin{algorithmic}[1]
        \REQUIRE $Crystals$ \# all samples in a batch
        \ENSURE \( \tilde{e}^a_{ij}, \tilde{e}^b_{ij}, \tilde{a}_{ijk} \)
        
        \STATE Initialize empty list ${\tilde{e}^a_{ij}}, {\tilde{e}^b_{ij}}, {\tilde{a}_{ijk}}$
        
        \FOR{each $g \in Crystals$}
            \STATE $L \leftarrow g.lattice$
            \STATE $I \leftarrow g.neighbor\_{image}$
            \STATE $r_{card} = g.r_{frac} @ L$    
            \STATE get $r_{i}, r_{j}$ from $r_{card}$
            \STATE $r_{j} = r_{j} + I @ L$
            \STATE $r_{ij} = r_i - r_j$
            \STATE $g\_\tilde{e}^a_{ij}, g\_\tilde{e}^b_{ij} = sRBF(r_{ij})$
            \STATE append $g\_\tilde{e}^a_{ij}$ to ${\tilde{e}^a_{ij}}$
            \STATE append $g\_\tilde{e}^b_{ij}$ to ${\tilde{e}^b_{ij}}$
            \IF{$angle\_nums \neq 0$}
                \STATE caculate $\theta_{ijk}$ from $r_{ij}$
                \STATE $g\_\tilde{a}_{ijk} = FT(\theta_{ijk})$
                \STATE append $g\_\tilde{a}_{ijk}$ to ${\tilde{a}_{ijk}}$
            \ENDIF
        \ENDFOR
        \STATE Concatenate ${\tilde{e}^a_{ij}}, \tilde{e}^b_{ij}, {\tilde{a}_{ijk}}$ along dimension 0

        \RETURN ${\tilde{e}^a_{ij}}, \tilde{e}^b_{ij}, {\tilde{a}_{ijk}}$
    \end{algorithmic}
\end{algorithm}

We improved the computational process to maximize GPU utilization. As illustrated in Alg.~\ref{alg: parallel process}, from lines 2 to 9, we first extract the lattice $L$, neighbor image $I$, and $r_{card}$ for each sample, storing this information in the lists $B\_r_{card}$, $B\_L$, and $B\_I$, respectively. Subsequently, we concatenate $B\_r_{card}$ and $B\_L$ along dimension 0, while $B\_I$ is assembled as a block diagonal matrix. From lines 12 to 19, we process all samples in one batch in parallel. By integrating all samples before performing the calculations, we significantly enhance GPU utilization and reduce CPU overhead.

\textbf{Redundancy Removal:} The reference CHGNet implementation introduces numerous tiny kernels that cannot saturate the GPU’s computing resource. There exists a lot of redundancy computations in the computation graph. 
For example, in sRBF calculation, the polynomial envelope function $u(r_{ij})$ contains redundant computations, as shown in Eq.~\ref{eq: sRBF coefficient naive}.
\begin{align*}
    u(r_{ij}) = & \ 1 - \frac{(p+1)(p+2)}{2} \left( \frac{r_{ij}}{r_{cut}} \right)^p \\
    & + p(p+2) \left( \frac{r_{ij}}{r_{cut}} \right)^{(p+1)} \\
    & - \frac{p(p+2)}{2} \left( \frac{r_{ij}}{r_{cut}} \right)^{(p+2)}
    \label{eq: sRBF coefficient naive}
    \tag{12}
\end{align*}

By factoring out common terms, we can eliminate these redundant calculations, as shown in Eq.~\ref{eq: sRBF coefficient optim}.
\begin{align*}
    u(r_{ij}) = \ 1 - \frac{p+2}{2}[(p+1)\xi^p +2p\xi^{(p+1)}-p\xi^{(p+2)}] 
    \tag{13}
    \label{eq: sRBF coefficient optim}
\end{align*}

where $\xi$ denotes $\frac{r_{ij}}{r_{cut}}$.

\begin{algorithm}[t]
    \caption{Parallel Computation of Basis in Batches}
    \label{alg: parallel process}
    \renewcommand{\algorithmicrequire}{\textbf{Input:}}
    \renewcommand{\algorithmicensure}{\textbf{Output:}}
    \begin{algorithmic}[1]
        \REQUIRE $Crystals$
        \ENSURE \(\tilde{e}^a_{ij}, \tilde{e}^b_{ij}, \tilde{a}_{ijk} \)
        \STATE Initialize empty list $B\_{r_{card}}, B\_L, B\_I$
        
        \FOR{each $g \in Crystals$}
            \STATE $L \leftarrow g.lattice$
            \STATE $I \leftarrow g.neighbor\_{image}$
            \STATE $r_{card} = g.r_{frac} @ L$
            \STATE append $I$ to $B\_I$
            \STATE append $L$ to $B\_L$
            \STATE append $r_{card}$ to $B\_{r_{card}}$
        \ENDFOR
        
        \STATE Concatenate $B\_{r_{card}}, B\_L$ along dimension 0
        \STATE Concatenate $B\_I$ as block diagonal matrix
        \STATE get $B\_r_{i}, B\_r_j$ from $B\_r_{card}$
        \STATE $B\_r_{j} = B\_r_{j} + B\_I @ B\_L$
        \STATE $B\_r_{ij} = B\_r_i - B\_r_j$
        \STATE $\tilde{e}^a_{ij}, \tilde{e}^b_{ij} = sRBF(B\_r_{ij})$
        \IF{$angle\_nums \neq 0$}
            \STATE caculate $B\_\theta_{ijk}$ from $B\_r_{ij}$
            \STATE $\tilde{a}_{ijk} = FT(B\_\theta_{ijk})$
        \ENDIF
        
        \RETURN ${\tilde{e}^a_{ij}}, \tilde{e}^b_{ij}, {\tilde{a}_{ijk}}$
    \end{algorithmic}
\end{algorithm}

\textbf{Computation Graph Reconstruction:} There are many small matrix multiplications (GEMMs) and they can be fused or packed together to increase the parallelism, as they share the same input or have no dependency on each other. For instance, a bundle of linear layers sharing the same input can be fused into a larger linear layer by weights concatenation, as depicted in Fig.~\ref{fig: Packing strategy}(a). In GatedMLP operation, the same computations such as layer normalization $LN()$, Sigmoid activations, etc are contained in two channels that can be merged, as shown in Fig.~\ref{fig: Packing strategy}(b). The $ sigmoid(x) $ calculation can be merged due to $silu(x) = x \cdot sigmoid(x)$. The $silu(x)$ result can be derived by multiplying $x (batch\left[0 \right])$, as the dashed line shows in Fig.~\ref{fig: Packing strategy}(b). Computation Graph Reconstruction is a critical technique in deep learning applications that combines multi-kernel into one efficient kernel, thereby reducing multiple kernel launches and data transfer overhead. 

\begin{figure}
    \centering
    \includegraphics[width=1.0\linewidth]{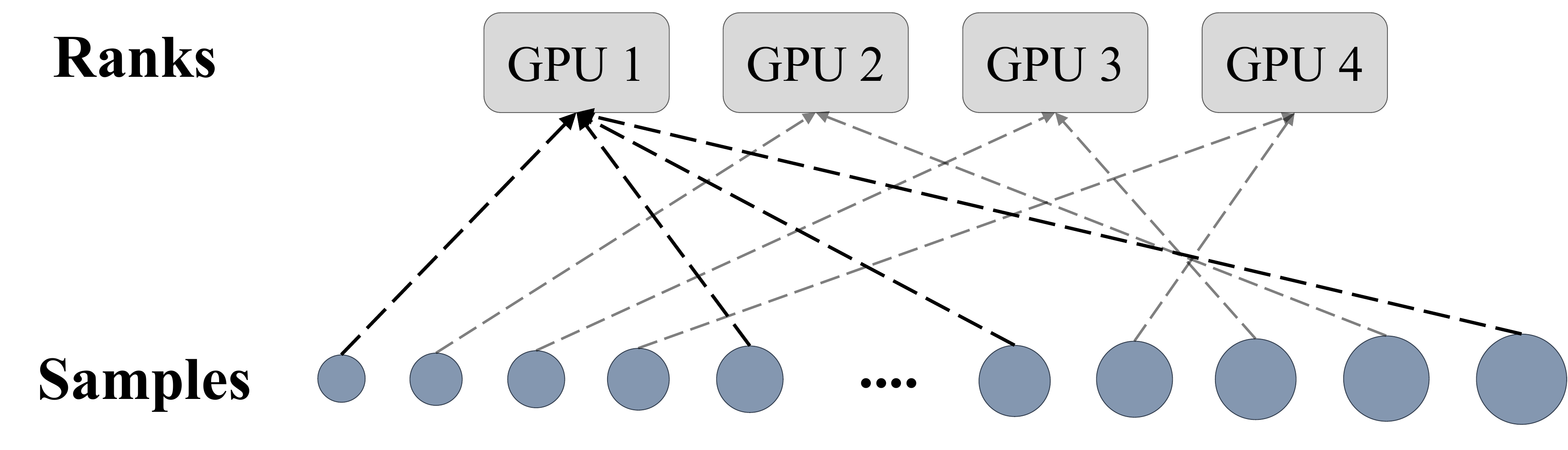}
    \caption{Load balance sampler. The strategy for distributing samples within a batch across multiple GPUs.}
    \label{fig: balance sampler}
\end{figure}

\begin{figure}
    \centering
    \includegraphics[width=1.0\linewidth]{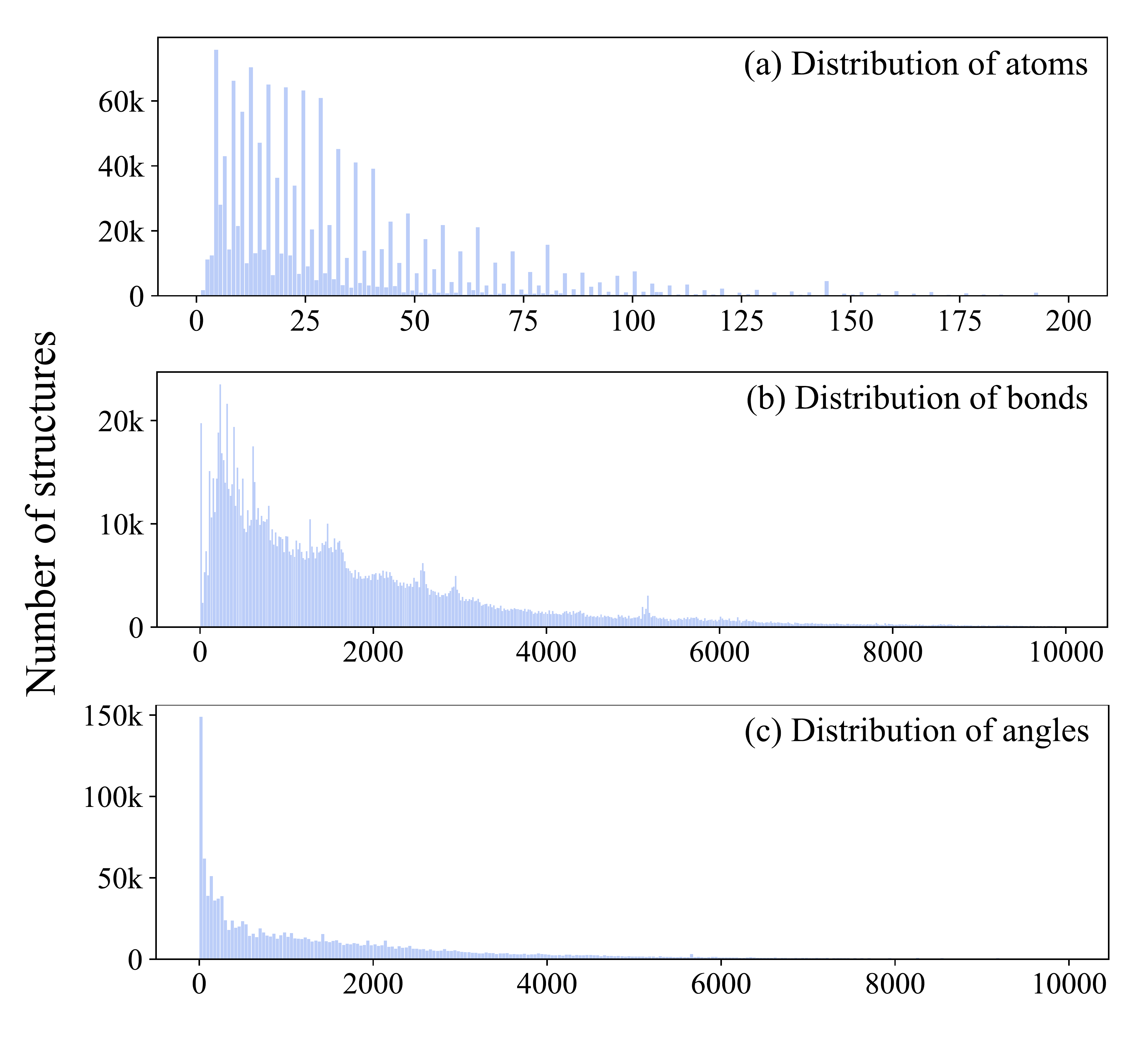}
    \caption{The atom/bond/angle distribution of MPtrj dataset.}
    \label{fig: distribution of MPtrj}
\end{figure}

\textbf{Load Balance Sampler:} The reference CHGNet implementation does not fully exploit the parallelism in the model’s computation graph. CHGNet can be trained only on a single GPU. We support multi-GPU training via data parallelism. As the multi-GPU training is employed, the batch size can be greatly increased. In this situation, the load imbalance problem is obvious in multi-GPU training between the samples in a batch as the number of atoms, bonds, and angles of different molecules varies substantially. This imbalance results in a large synchronization overhead in the multi-GPU setting. Fig.~\ref{fig: distribution of MPtrj} describes the distribution of training samples, including the number of nodes, bonds, and angles and their frequency in the MPTrj dataset. It can be easily recognized that the frequency follows a long-tail distribution. To solve the load imbalance problem, we first calculate the total number of atoms, bonds, and angles for each sample in the global batch and sort them in ascending order. 
Then, each GPU selects the smallest and largest samples from the remaining samples in turn, until all samples are allocated. When the number of GPUs is 4, the sample allocation scheme is shown in Fig.~\ref{fig: balance sampler}. This approach ensures that the computational load for each instance is as balanced as possible.

\textbf{Learning Rate Schedule:} When training with large batch sizes, an excessively low learning rate results in insufficient parameter updates, causing the model to move too slowly in the parameter space and ultimately affecting its accuracy. To fully leverage the advantages of large batches, we employ an adaptive learning rate adjustment strategy as follows: 
\begin{equation*}
	\begin{split}
        init_{LR} = \frac{batch size}{k} \times 0.0003
    \end{split}
    \tag{14}
    \label{eq: lr adjust strategy}
\end{equation*}
where $k$ is a hyper-parameter and we set $k$ as 128. This approach adjusts the learning rate in proportion to the batch size, ensuring a steady and reliable convergence.

\textbf{Other Optimization:} 

\textit{Data Prefetch:} This technique is employed to enhance the efficiency of batches loading during the training process. While the current batch is being processed, data prefetch asynchronously transfers the next mini-batch from CPU memory to GPU memory, reducing data wait times and optimizing computational resource usage. Specifically, this strategy utilizes separate streams for data transfer, allowing the copy operation to occur concurrently with the forward and backward passes. 

\textit{Communication Overlap:} After each backward pass, a global all-reduce operation is required to obtain the gradients. This communication overhead is unavoidable. However, instead of waiting for all gradients calculations to finish,  we can perform all-reduce once after the gradient calculation of a part of parameters is completed while the other part of gradients are being calculated. This overlap of communication and calculation minimizes idle time for each compute node and accelerates the overall training process. 

\section{Experiment Setup}

\textit{Hardware and software stacks:} 
All numerical tests are conducted using the GPU cluster. Each node is equipped with two 64-core Intel Xeon Platinum 8358 CPUs and 8 NVIDIA A100 GPUs. Each node has 1TB host memory and each A100 GPU has a memory capacity of 80GB. The CPU bandwidth is 11.2 GT/s and the GPU bandwidth is 1935 GB/s. Node communication operates in a non-blocking fat-tree topology. GCC 11.3.0 is chosen for compiling CPU code and CUDA 12.2 is our GPU compiler. The code is developed on the PyTorch 2.3.1 deep learning platform. The packages pymatgen, ase, etc are also used.

\textit{Dataset:} 
The MPtrj dataset consists of 1,580,395 inorganic crystal structures composed of 89 elements, with data including 1,580,395 energies, 7,944,833 magnetic moments, 49,295,660 forces, and 14,223,555 stresses, all calculated using DFT. These structures and labels are extracted from both static and relaxation trajectories obtained through GGA/GGA+U calculations in the Materials Project. In constructing the atom and bond graph, we set the default cutoff distances to 6 \text{\AA} and 3 \text{\AA}, respectively. For training FastCHGNet, the dataset is divided into training, validation, and test sets in a 0.9:0.05:0.05 ratio.

\textit{Parameters Setting:} 
FastCHGNet has 429,046 trainable parameters. The radial and angular basis number is set to 31. The atom, bond, and angle features are embedded into 64-dimensional vectors. The smoothing coefficient $p$ is set to 8. The model predicts energy, force, stress, and magnetic moment. The loss function in backpropagation is Huber loss, with the prefactor defined as 2, 1.5, 0.1, and 0.1  respectively. `Adam' optimizer is adopted. The initial learning rate is 0.0003 and the cosine annealing scheduler is applied. The activation functions are Sigmoid and SiLU. We train the model for 30 epochs with a batch size of 128. The reference CHGNet is available at: 
\url{https://github.com/CederGroupHub/chgnet/tree/main/chgnet/pretrained/0.3.0}, v0.3.0.

\section{Evaluation}
\subsection{Convergence results}
Table.~\ref{tab: model accracy} describes the MAE result (the lower the better) of CHGNet and FastCHGNet on the MPtrj testing set. In Energy prediction, FastCHGNet reaches a higher accuracy. Overall, FastCHGNet demonstrates comparable accuracy compared to CHGNet in predicting energy(meV/atom), force(meV/$\mathring{\text{A}}$), stress(GPa) and magmom($\mu_B$). FastCHGNet(version `w/o head') means the output layer is not decoupled. This version achieves a higher precision on the four predictions than the reference CHGNet. This is a result of using larger batch sizes and fine-tuning the learning rate. The parameter of the FastCHGNet(version `w/o head') is slightly reduced because some unnecessary modules are removed. FastCHGNet(version `F/S head') means the output layer is decoupled by Force head and Stress head. Thereby the number of parameters is larger than the reference CHGNet. The Force and Stress decomposition modules may result in a small decrease in the precision of force and pressure.  
\begin{table*}[h!]
    \centering
    \caption{The Mean Absolute Error(MAE) of CHGNet, FastCHGNet on MPTrj test dataset.}
    \label{tab: model accracy}
    \begin{tabular}{ccccccc}
    \toprule
        \textbf{model} & \textbf{version} &\textbf{param} & \textbf{Energy(meV/atom)} & \textbf{Force(meV/$\text{\AA}$)} & \textbf{Stress(GPa)} & \textbf{Magmom(\textbf{$m\mu_B$)}} \\ 
        \midrule
        CHGNet & v0.3.0 & 412.5K & 29 & 68 & 0.314 & 37 \\ 
        \cline{1-7}
        FastCHGNet & w/o head & 411.2k & 26 & 62 & 0.270 & 35 \\
        \cline{1-7}
        FastCHGNet & F/S head & 429.1K & 16 & 73 & 0.479 & 36 \\
    \bottomrule
    \end{tabular}
\end{table*}

\begin{figure}
    \centering
    \includegraphics[width=1.0\linewidth]{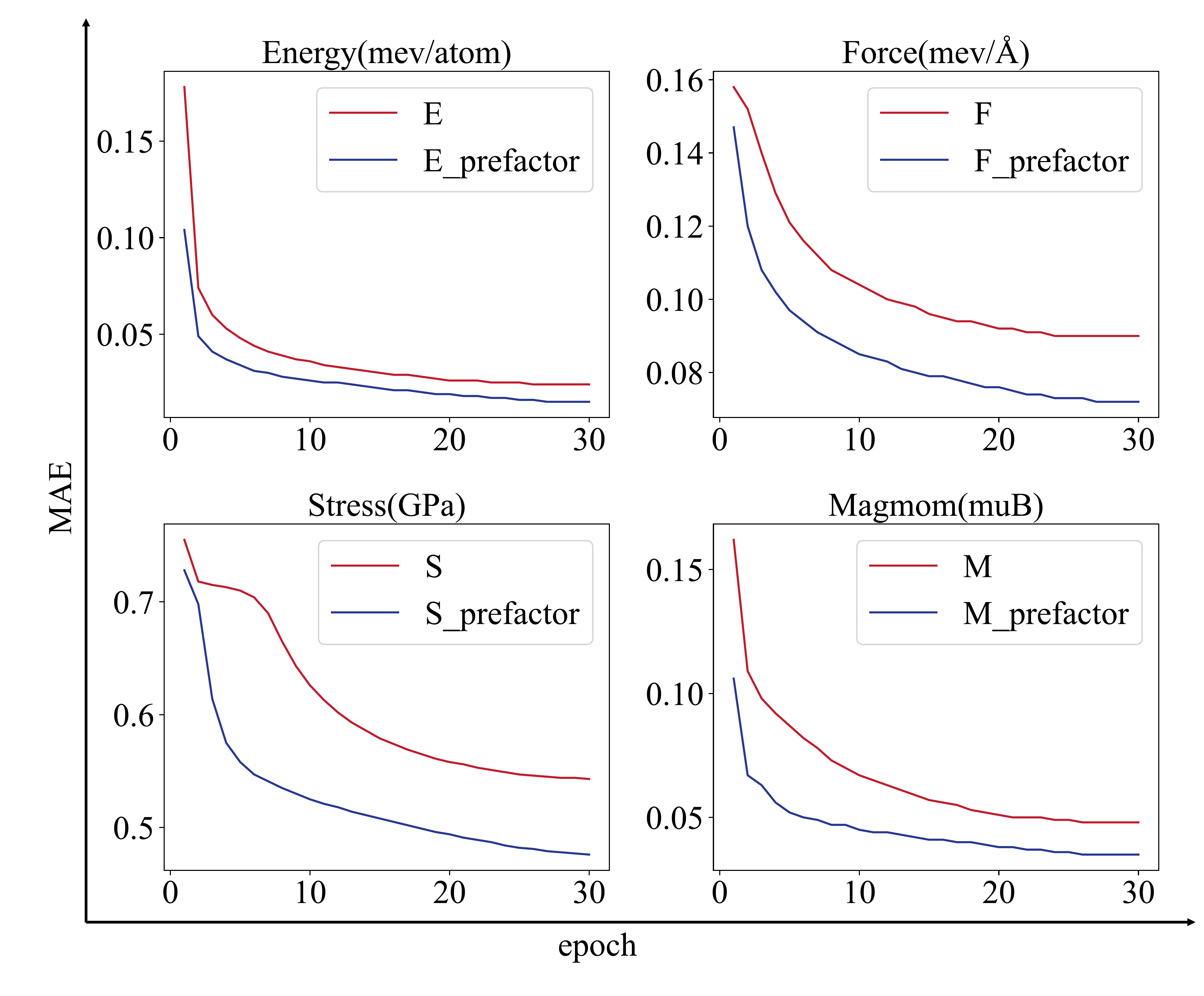}
    \caption{The convergence of fine-tuned learning rate in terms of Energy, Force, Stress, and Magmom. The red line means the default learning rate. The blue line means the learning rate is multiplied by a prefactor.}
    \label{fig: factor LR schedule}
\end{figure}

Fig.~\ref{fig: factor LR schedule} illustrates the effectiveness of the tuning of the learning rate. We increase the batch size to 2048 and trained for 30 epochs, investigating the convergence of Energy (E), Force (F), Stress (S), and Magnetic Moment (M) where MAE is the evaluation metric. The red curves represent the results using the default learning rate (0.003), where E, F, S, and M converge to 24 meV/atom, 90 meV/$\text{\AA}$, 0.543 GPa, and 48 $m\mu_B$, respectively. The blue curves show the results after adjusting the learning rate according to Eq.~\ref{eq: lr adjust strategy}, achieving an improved accuracy of 15 meV/atom, 72 meV/$\text{\AA}$, 0.476 GPa, and 35 $m\mu_B$, respectively.

\begin{figure}[h]
\centering
    \includegraphics[width=0.45\textwidth]{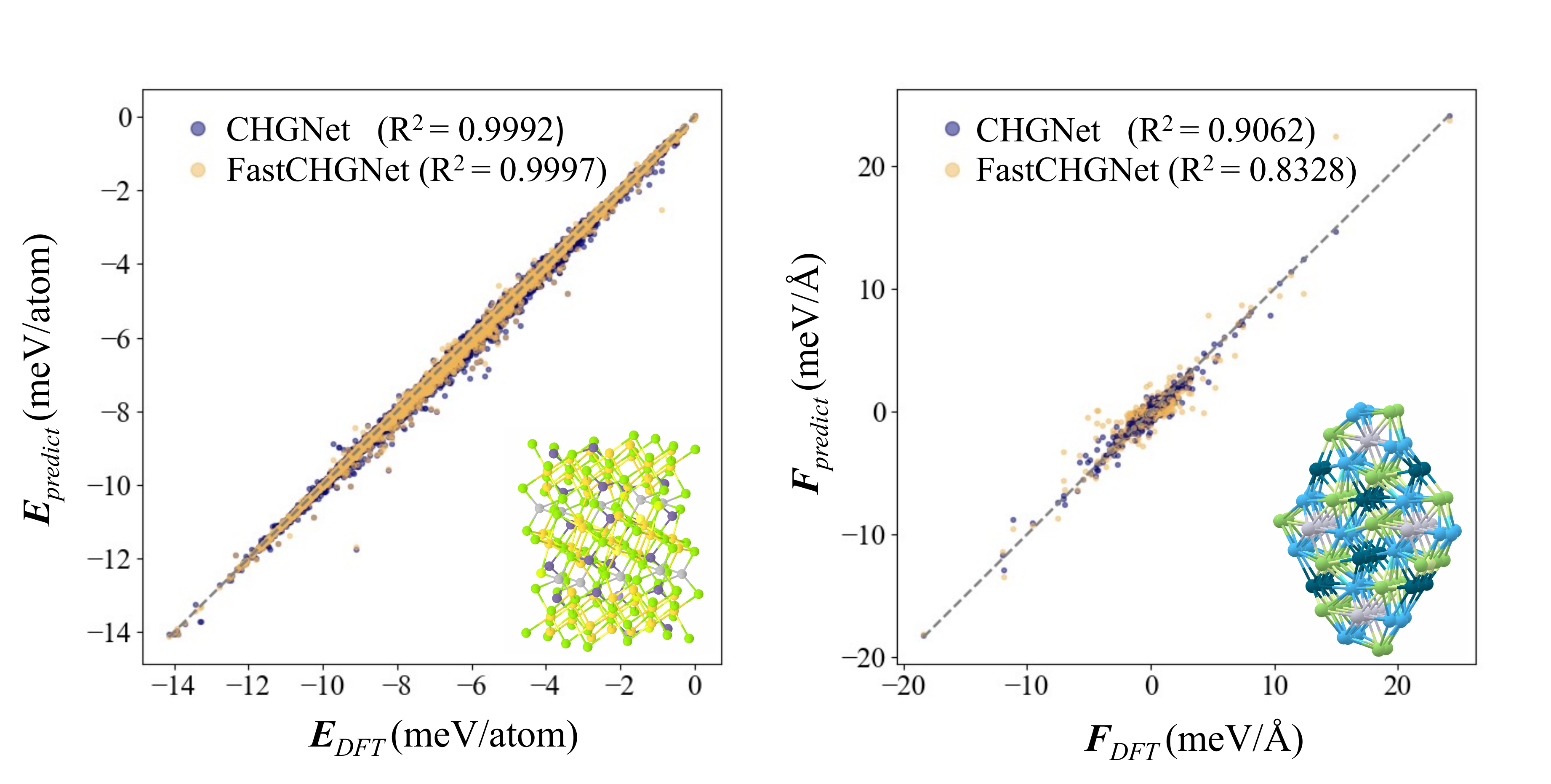}
    \caption{CHGNet and FastCHGNet performance contrasted with DFT results on the testing dataset.}
    \label{fig: predict and dft}
\end{figure}

Fig.~\ref{fig: predict and dft} shows how far the fitting results of CHGNet and FastCHGNet in the testing set deviate from those of DFT (ground truth) in terms of energy and force. The x-axis represents the DFT results, and the y-axis represents the predicted results. The testing snapshots would fall on the solid black line when the predictions of the neural network force field match exactly the DFT results. A more accurate CHGNet/FastCHGNet model is one where the predictions are as close as possible to DFT results, with the data points lying as close as possible to the solid black line. The R² is calculated and an R² value closer to 1 indicates a better fit of the model to the data. FastCHGNet has a higher R² than CHGNet in energy but a lower R² for force. We randomly selected four systems in the testing dataset and their visualizations are in the lower right corner of each subplot in Fig.~\ref{fig: predict and dft}.

\subsection{Single GPU}
\begin{figure}
    \centering
    \includegraphics[width=\linewidth]{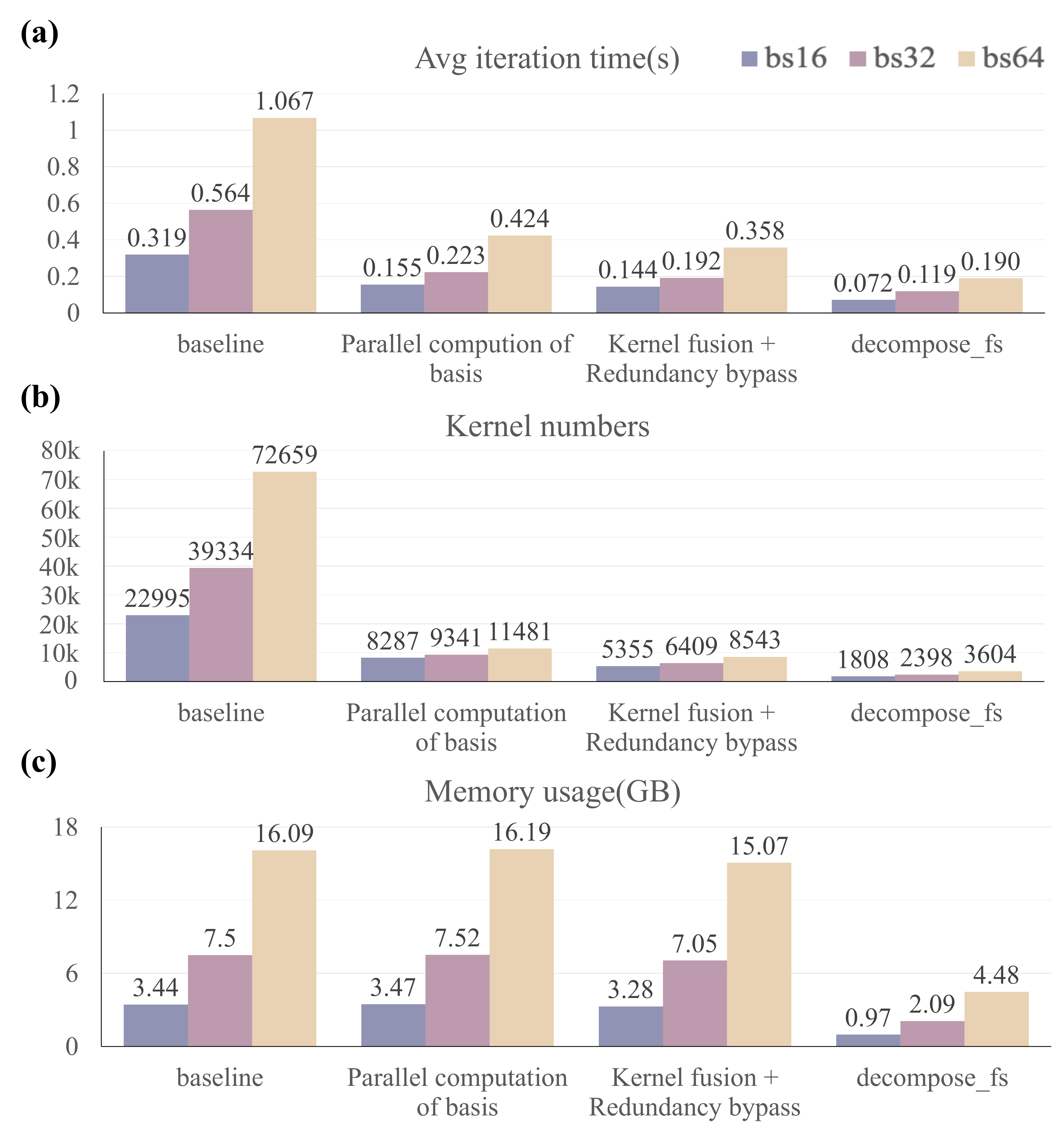}
    \caption{The average iteration time, the number of launched kernels, and the memory usage under step-by-step system optimization.}
    \label{fig: step by step optim}
\end{figure}

\textit{Iteration time \& Kernel number:} We compare the training performance of CHGNet and FastCHGNet across different batch sizes (16/32/64) on a single NVIDIA Tesla A100 GPU. As shown in Fig.~\ref{fig: step by step optim}(a,b), after a series of optimizations, FastCHGNet achieves a $4.43–5.62\times$ reduction in training time and a $12.72–20.16\times$ decrease in the number of launched kernels. In CHGNet,  operations such as `sRBF' are serial processed and, thus have relatively low computational and resource usage. CHGNet fails to fully leverage the parallelism and resources of GPUs. Therefore, we modified a series of serial computations into parallel computations. And this strategy(`Parallel computation of basis') results in a $2.06–2.52\times$ speedup. This improvement is primarily due to the conversion of a series of serial calculations into parallel operations. This approach offers two key benefits: first, alleviating the CPU overhead. FastCHGNet reduces the number of launched kernels from 72659 to 11481 (batch size=64); second, by rewriting the inefficient calculations to an efficient parallel mode, we can enhance GPU utilization. We also employ the `Kernel fusion + Redundancy bypass' strategy to further reduce the number of launched kernels, resulting in an additional $1.08–1.18\times$ speedup. Furthermore, we implement the decoupling of energy-force and energy-stress calculations by directly computing forces and stresses. The `decoupling' strategy does not need to compute second-order derivatives in weights updating, resulting in an additional $1.88–2\times$ speedup.

\textit{Memory usage:} We also compared the GPU memory usage of CHGNet and FastCHGNet under different batch sizes, as shown in Fig.~\ref{fig: step by step optim}(c). GPU memory will slightly increase when `Parallel computation of basis' is applied. We concatenate certain features to facilitate parallel processing on the GPU, while some features (such as the offset vector) require padding with zeros, leading to increased memory demands. By employing kernel fusion and redundancy removal, we eliminate redundant computations and modules in CHGNet, resulting in a memory reduction of $1.05–1.07\times$. Furthermore, when we decouple force and stress, the memory usage decreases by a factor of $3.38–3.50\times$, which is attributed to FastCHGNet's ability to compute force and stress without relying on first-order derivatives, thereby eliminating the need to store intermediate values from the first-order derivative computations.

\subsection{Multi GPUs}
\begin{figure}
    \centering
    \includegraphics[width=1.0\linewidth]{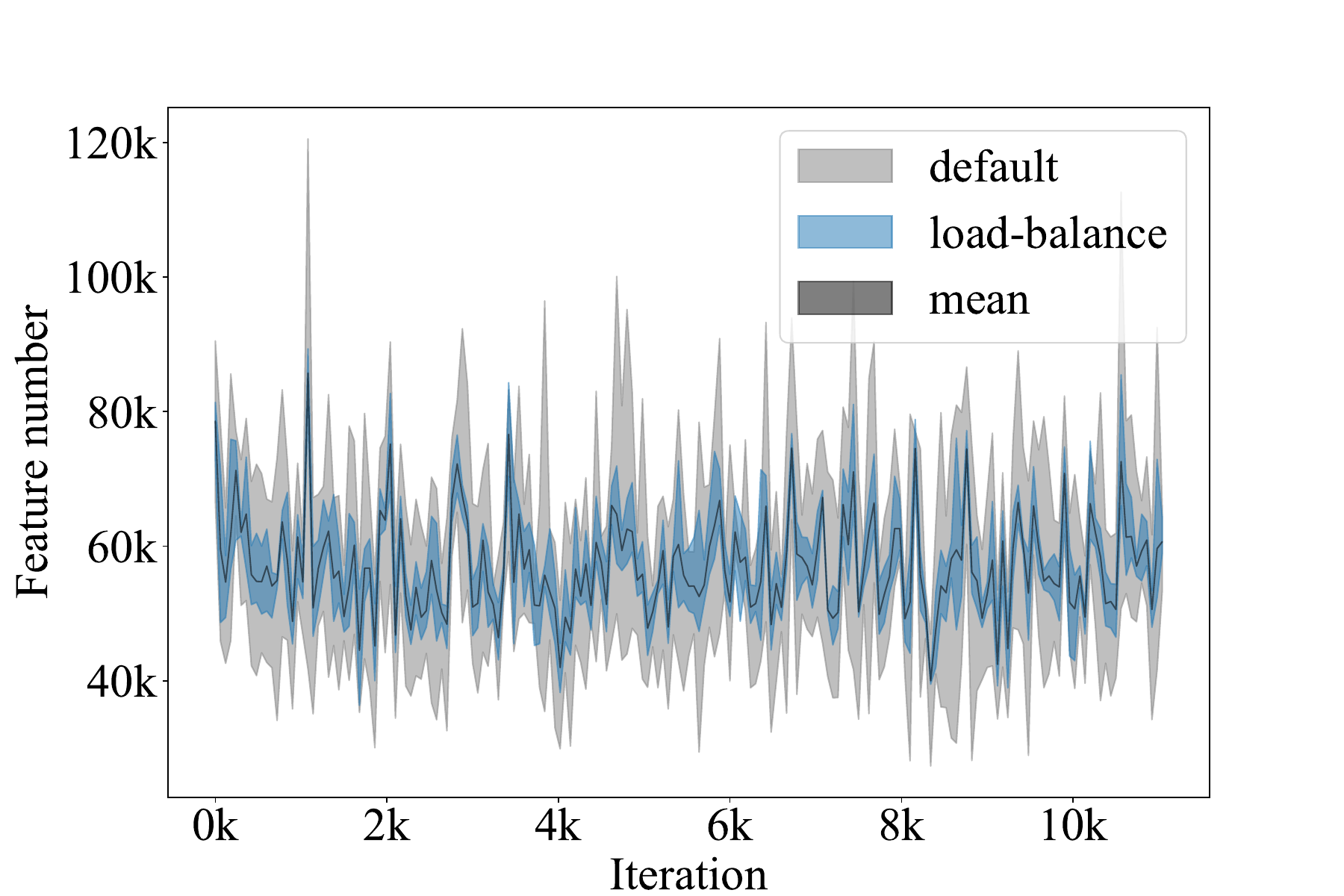}
    \caption{The feature number of default sampler and load balance sampler.}
    \label{fig: load balance}
\end{figure}

\textit{Load Balance:} Training on a single A100 GPU still requires 35.4 hours although the optimizations above have significantly improved training speed compared to the reference CHGNet. Training by a larger batch size with multi-GPUs can reduce the absolute training time. With an increase in training batch size, the coefficient of variance(a criterion used to describe load imbalance; the higher the variance, the more severe the imbalance) is 0.186(with the default mini-batch size of 32 on 4 GPUs). This indicates that the number of nodes, bonds, and angles varies significantly, as shown in the gray-shaded area of Fig.~\ref{fig: load balance}. The x-axis of Fig.~\ref{fig: load balance} is iteration and the y-axis(Feature number) denotes the workload of the training batch. The feature number is the summation of atom number, bond number, and angular number. We design a sampler to address the load imbalance issue, depicted in Section III. By assigning the largest and smallest samples to the same GPU, the coefficient of variance reduces to 0.064 after this strategy, as indicated by the blue-shaded area in Fig.~\ref{fig: load balance}.

\begin{figure}
    \centering
    \includegraphics[width=1.0\linewidth]{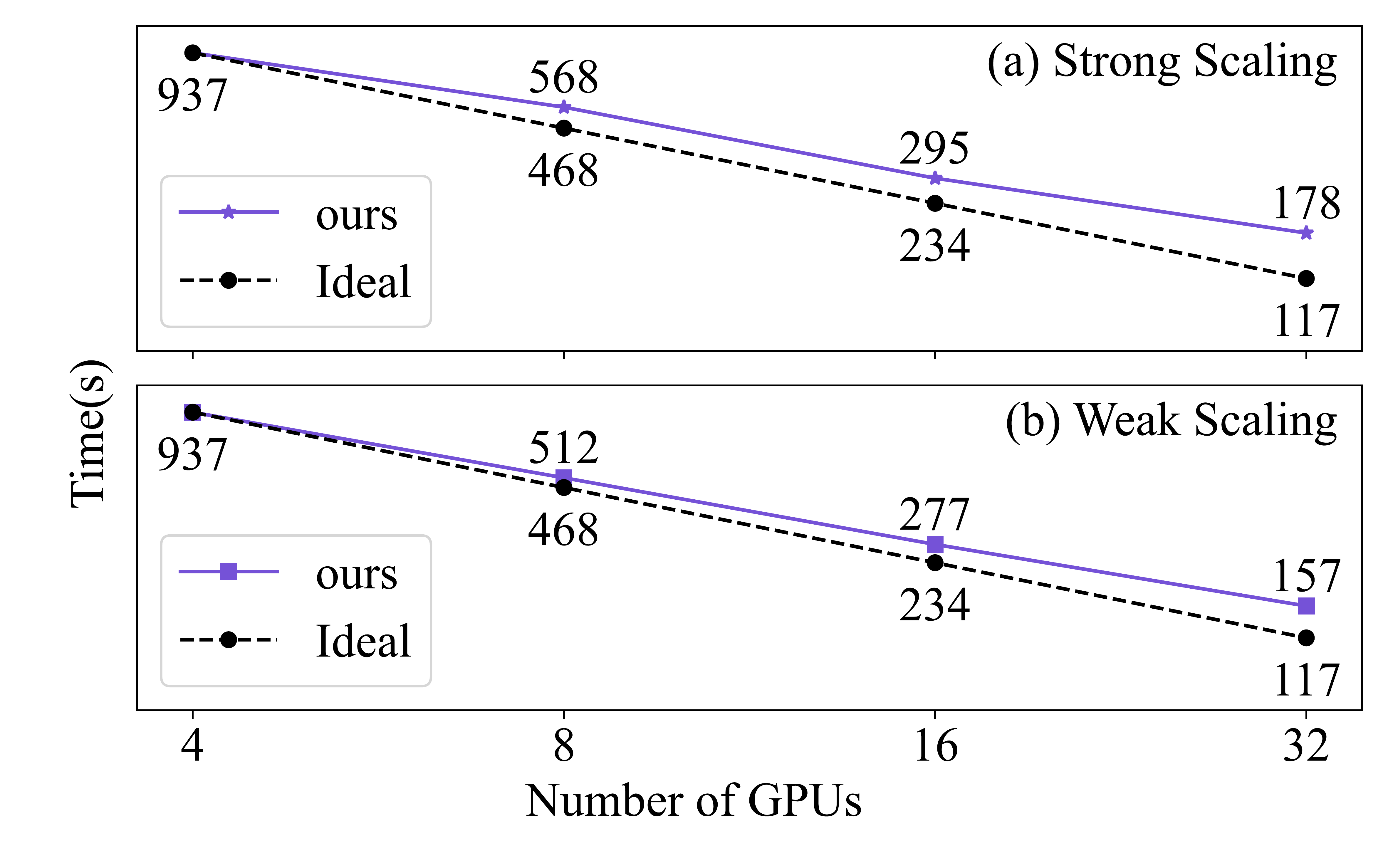}
    \caption{The strong scaling and weak scaling of FastCHGNet in 4, 8, 16, 32 GPUs.}
    \label{fig: strong/weak scaling}
\end{figure}

\textit{Strong Scaling:} To evaluate the strong scalability of FastCHGNet, we conduct experiments using 4 to 32 GPUs with a global batch size set to 2048. Fig.~\ref{fig: strong/weak scaling}(a) illustrates the strong scalability results. The horizontal axis represents the number of GPUs utilized (note: we use only 4 GPUs for per compute node), while the vertical axis indicates the time required to train one epoch. The speedup is $1.65\times$ when 8 GPUs are utilized with the efficiency $82.5\%$. When trained by 16 GPUs, we get a $3.18\times$ speedup compared with training by 4 GPUs. The scaling efficiency is $79.5\%$. The speedup is attributed to the reduced computational workload per GPU. This is because, as the number of GPUs increases, the mini-batch size per GPU decreases.
When the training GPUs increase to 32, the training time is reduced by $5.26\times$ compared to training with 4 GPUs. The scaling efficiency is $66\%$. The lower scaling efficiency comes from the significant communication overhead that arises as the number of GPUs increases.

\textit{Weak Scaling:} In the weak scaling tests of FastCHGNet, we set the mini-batch size to 512. Testing is performed on 4, 8, 16, and 32 GPUs. The scaling performance is shown in Fig.~\ref{fig: strong/weak scaling}(b). The x-axis of Fig.~\ref{fig: strong/weak scaling}(b) denotes the number of GPUs and the y-axis is the training time of one epoch. The scaling efficiencies for 4, 8, 16, and 32 GPUs are $91.5\%$, $84.6\%$, and $74.6\%$, respectively.

\subsection{Molecular dynamics in real applications}
We also compared the performance of CHGNet and FastCHGNet in molecular dynamics. The results are shown in Table.~\ref{tab: Inference time}. We randomly select three systems(LiMnO2, LiTiPO5, Li9Co7O16). They are in different sizes, with feature numbers (the summation of the atoms, bonds, and angles number) of 1088, 3582, and 10188, respectively. The inference speed of FastCHGNet is 2.63 to 3.03 times faster than that of CHGNet. To be more specific, one-step molecular dynamics time before and after optimization is denoted as $t$ and $t^*$. The speedup $r=\frac{t}{t^*}$ is 2.86, 2.63, and 3.03. The speedup is not as high as the training process. This is due to the given structure is processed step by step in molecular dynamics. One structure molecular dynamics has a low computational cost. In this situation, the GPU computational power cannot be fully utilized. 
\begin{table*}
    \centering
    \caption{The time required for one-step molecule dynamics of CHGNet, FastCHGNet on LiMnO2, LiTiPO5, Li9Co7O16 structures.}
    \label{tab: Inference time}
    \begin{tabular}{ccccccc}
    \toprule
        \textbf{crystal} & \textbf{atoms} &\textbf{bonds} &\textbf{angles} &\textbf{CHGNet} & \textbf{FastCHGNet} &\textbf{speedup} \\ 
        \midrule
        LiMnO2 & 8 & 336 & 744 & 0.022 & 0.0077 &2.86 \\ 
        \cline{1-7}
        LiTiPO5 & 32 & 1258 & 2292 & 0.021 & 0.0076 &2.63\\
        \cline{1-7}
        Li9Co7O16 & 32 & 1780 & 8376 & 0.023 & 0.0077 &3.03\\
    \bottomrule
    \end{tabular}
\end{table*}

\section{Related work}
\textbf{Dedicated ML potentials:} Machine Learning(ML) methods have been developed to accurately predict atomistic potential energy surfaces (PES) for various applications. There are two main categories of ML potentials. The first category is dedicated ML potentials, which are specifically crafted to describe the PES of a particular system or a very limited class of systems. They require DFT calculations before training, making them impractical for large-scale applications because DFT calculations are expensive~\cite{smith2017ani, devereux2020extending}. On the other hand, the particular system or the limited class of systems has a relatively low number of degrees of freedom. Thus, dedicated ML potentials are easy to train to converge. The category models tend to be small. The DeePMD~\cite{wang2018deepmd}, BPNN~\cite{behler2007generalized}, EANN~\cite{zhang2019embedded}, PAINN~\cite{schutt2021equivariant}, NequIP~\cite{batzner20223}, NewtonNet~\cite{haghighatlari2022newtonnet}, GemNet~\cite{gasteiger2021gemnet}, SpookyNet~\cite{unke2021spookynet}, DimeNet++~\cite{gasteiger2020fast} are dedicated ML potentials.

\textbf{Universal ML potentials:} The second category is universal ML potentials which are developed to simulate entire classes of molecules or crystals. Once universal ML potentials are trained, they can be applied across a broad range of systems without further DFT calculations being added. Therefore, universal ML potentials can achieve linear scaling in physical structures. A promising universal model needs a wide range of high-accuracy training data(to provide enough data diversity) and a well-designed model(to describe the potential energy surface). The SevenNet~\cite{park2024scalable}, MACE~\cite{batatia2022mace, batatia2023foundation}, CHGNet~\cite{deng2023chgnet}, M3GNet~\cite{chen2022universal}, ALIGNN~\cite{choudhary2021atomistic}, MEGNet~\cite{chen2019graph}, CGCNN~\cite{xie2018crystal}, GPTFF~\cite{XIE2024} are universal ML potentials. In this paper, we focus on universal ML potentials. Among these universal ML potentials, CHGNet is the only charge-informed universal ML potential. Note that charge information can provide insights into ionic systems with electronic degrees. Most of the universal ML potentials are GNN backbone for the reason that the message passing scheme is an effective method for learning atomistic systems. 
 
\textbf{Neural network optimization:} Neural network optimization aims to enhance training and inference efficiency, scalability, and performance. Recently, various approaches have been proposed to address the challenges associated with deep neural networks. Among these approaches, deep learning compiler is one of the most prominent tools for improving model efficiency~\cite{ashouri2018survey}, such as TVM~\cite{chen18and}, TensorRT~\cite{zhang2022cdnet}, Ansor~\cite{zheng2020ansor}, XLA~\cite{snider2023operator}, etc. They can generate optimized codes automatically. However, they aim at making inferences more efficiently except for XLA. XLA can be used in model training while only supporting Pytorch~\cite{paszke2019pytorch} Framework. In this paper, we focus on model training.
Quantization is a popular neural network optimization strategy and has received significant attention recently~\cite{tailor2020degree, wang2022qgtc, ding2021vq, feng2020sgquant}. Quantization assigns different bits to different tensors to reduce memory consumption and computation overhead. GNN models can achieve faster inference through quantization. While in the training of machine learning interatomic potentials, quantization has not been successfully applied. The interatomic potential training is sensitive and has extremely high accuracy demand. DeePMD is trained using Float64 and other atomic potential models are trained by Float32. To the best of our knowledge, there are currently no atomic potential models that have been trained using half-precision. Although quantization has been well studied for CNNs, GNNs, and language models, there remains relatively little work utilizing quantization techniques in training atomic potential models. 
By designing an innovative optimizer, RLEKF, the training process of interatomic potentials can be accelerated\cite{hu2023AAAI}. RLEKF converges 8 to 10 times faster\cite{RJXB20250115006} than first-order methods on classical networks such as MPT\cite{Novikov_2021}, SNAP\cite{THOMPSON2015316}, etc. The parallel quasi-Newton optimizer, FastEKF, demonstrates significant training acceleration for neural network potentials, reducing the training time of DeePMD from days to minutes\cite{hu2024PPoPP}. EKF-based optimizers provide insights into training MLP-based neural network potentials, while their application to GNN-based potentials requires further exploration.
In summary, the training of GNN-based universal ML potentials is still a challenge. 

\section{Conclusion}
CHGNet is the state-of-the-art GNN-UIP model for charge-informed MD simulations, yet efficiently training this model poses a significant challenge. In this paper, we introduce FastCHGNet, an optimized implementation of CHGNet. The Force and Stress are fitted by the novel Force and Stress head. The memory requirement is greatly reduced. To fully utilize GPU computation resources, a lot of strategies such as kernel fusion, redundancy bypass, etc, have been used to enhance GPU utilization. We also propose a Load Balance Sampler to ensure a relatively even distribution of the computational workload across GPUs. The training time of the reference CHGNet is 8.3 days. Without sacrificing accuracy, the training time of FastCHGNet(without force/stress decoupling) can reduce to 3.79 hours. The training time of a more aggressive version (force/stress will be decoupled) can be reduced to 1.53 hours.


In the future, we plan to design more lightweight modules to replace the currently time-consuming operations, while maintaining model accuracy. In the meantime, we will try to apply model compression and quantization to further accelerate the training process of universal interatomic potentials.

\section*{Acknowledgment}
This work is supported by the following funding: National Science Foundation of China (92270206, T2125013, 62372435, 62032023, 61972377,  61972380, T2293702), China National Postdoctoral Program for Innovative Talents (BX20240383), CAS Project for Young Scientists in Basic Research (YSBR-005), the Innovation Funding of ICT, CAS under Grant No. E463030. The authors thank the ICT operations team for their strong support. 

\bibliographystyle{IEEEtran}
\bibliography{ref}
\end{document}